\documentclass[runningheads]{llncs}
\usepackage{hyperref}
\usepackage{subcaption}
\usepackage{tabularx}
\usepackage{float}
\usepackage[font=scriptsize]{caption}
\usepackage{pdfpages}
\hypersetup{
	colorlinks=true,
	linkcolor={red!50!black},
	citecolor={blue!60!black},
	urlcolor={blue!80!black},
	pdfauthor={Sadeghibogar Zahra, Berti Alessandro, Pegoraro Marco, van der Aalst Wil M.P.},
	pdftitle={{Process Mining in HPC Workflows}},
	pdfsubject={Process Mining and High Performance Computing},
	pdfkeywords={High Performance Computing, SLURM, Scientific workflow, Process mining},
	pdfproducer={LaTeX},
	pdfcreator={pdfLaTeX},
	bookmarksopen=true
}

\begin{document}
\title{Applying Process Mining on Scientific Workflows: a Case Study on High Performance Computing Data\thanks{The authors gratefully acknowledge the German Federal Ministry of Education and Research (BMBF) and the Ministry of Education and Research of North-Rhine Westphalia for supporting this work/project as part of the NHR funding. Also, we thank the Alexander von Humboldt (AvH) Stiftung for supporting our research.}}
\titlerunning{Applying Process Mining on HPC Scientific Workflows}

\author{
Zahra Sadeghibogar\orcidID{0000-0002-6340-9669},
Alessandro Berti\orcidID{0000-0002-3279-4795},
Marco Pegoraro\orcidID{0000-0002-8997-7517},
Wil M.P. van der Aalst\orcidID{0000-0002-0955-6940}
}
\authorrunning{Zahra Sadeghibogar et al.}
\institute{Chair of Process and Data Science, RWTH Aachen University, Aachen, Germany
\email{\{sadeghi, a.berti, pegoraro, wvdaalst\}@pads.rwth-aachen.de }}

\maketitle

\begin{abstract}
Computer-based scientific experiments are becoming increasingly data-intensive, necessitating the use of High-Performance Computing (HPC) clusters to handle large scientific workflows. These workflows result in complex data and control flows within the system, making analysis challenging. This paper focuses on the extraction of case IDs from SLURM-based HPC cluster logs, a crucial step for applying mainstream process mining techniques. The core contribution is the development of methods to correlate jobs in the system, whether their interdependencies are explicitly specified or not. We present our log extraction and correlation techniques, supported by experiments that validate our approach, enabling comprehensive documentation of workflows and identification of performance bottlenecks.

\end{abstract}

\section{Introduction}
\label{sec:introduction}
\begin{figure}[t]
\centering
\centering
\includegraphics[width=0.85\textwidth,page=1]{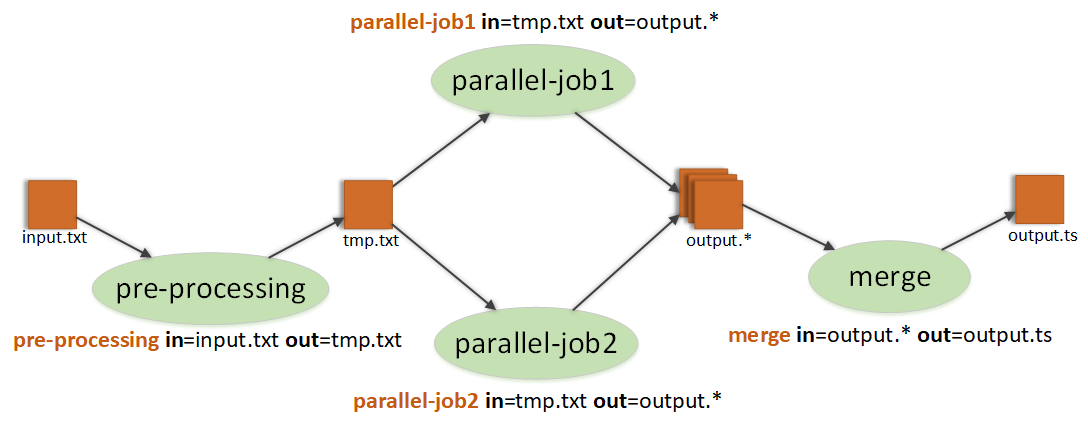}
\caption{Illustration of a Scientific Workflow: Starting with a single input file, the process involves pre-processing, parallel job execution, and merging outputs into a single final file.} \label{scientific_workflow}
\end{figure}

A \emph{workflow} is a description and automation of a process, in which data is processed by different logical data processing activities according to a set of rules. A \emph{scientific workflow} is an ensemble of scientific experiments, described in terms of scientific activities with data dependencies between them~\cite{DBLP:journals/fgcs/DeelmanGST09}.  Scientific workflows allow scientists to model and express the entirety of data processing steps and their dependencies. Fig.~\ref{scientific_workflow} shows an example of a scientific workflow depicted as a flow chart, where each task is associated with a command.

Scientific workflows, characterized by their massive data processing needs, are often automated and necessitate parallel processing on platforms such as cloud or HPC clusters~\cite{DBLP:journals/fgcs/DeelmanGST09,DBLP:conf/services/ZengHA11}. In this context, process mining emerges as an invaluable tool for workflow comprehension and detection of possible optimizations.

Over the past decades, there has been a growing interest in the field of process mining~\cite{DBLP:books/sp/22/PMH2022}. Process mining aims to extract information about processes from event logs, i.e., execution histories. This paper applies process mining to existing scientific workflows with the following goals:

\begin{itemize}
\item \emph{Documentation of scientific workflows}: reporting which commands are executed and in which order. We pursue this goal by using process discovery techniques, one of the main branches of process mining~\cite{DBLP:books/sp/22/PMH2022}. Process discovery techniques assume that every record in the event log contains at least: (i) a reference to the executed activity, (ii) a reference to the identifier that associates an event with a particular execution of the process, and (iii) the timestamp at which the event occurred.
\item \emph{Detection of bottlenecks affecting the execution of scientific workflows}: We enrich the process model discovered in the previous step with the obtained performance results.
\end{itemize}

While the techniques proposed in this paper can be applied to any workflow system, we focus on the SLURM system to promote applicability. SLURM is a common choice for workflow management in HPC clusters, governing the RWTH HPC cluster\footnote{\url{https://help.itc.rwth-aachen.de/service/rhr4fjjutttf/}}, one of the most widely used platforms in the field.

\begin{figure}[t]
\centering
\includegraphics[width=\textwidth,page=1]{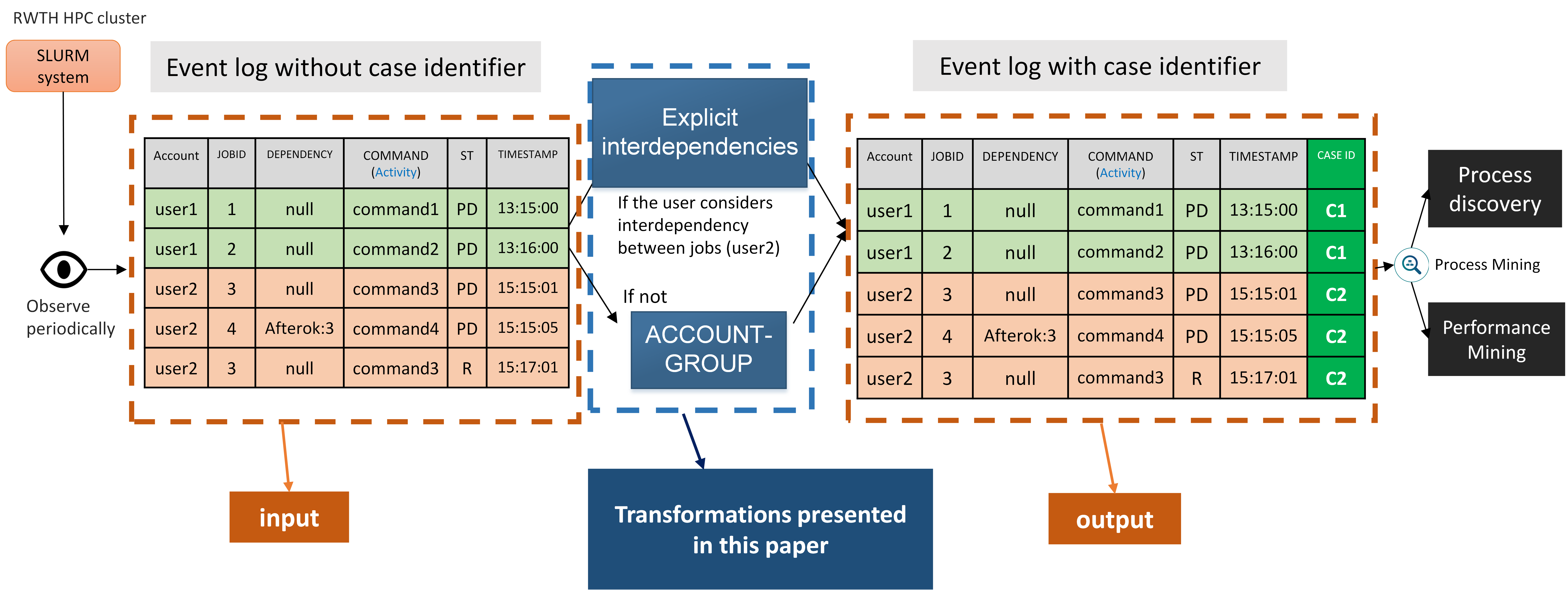}
\caption{An overall view of the proposed approach} \label{overall}
\end{figure}

The issue in examining the logs obtained from a given workflow system is the absence of a clearly defined case identifier that groups events associated with the same execution. To apply process mining to these logs, it is necessary to study the correlation between tasks that are running on the HPC cluster. Fig.~\ref{overall} shows an overall view of our approach. The RWTH HPC cluster is observed periodically, and an input log is generated. Based on how users execute their jobs on SLURM~\cite{DBLP:conf/jsspp/YooJG03}, we propose two different approaches to assign case IDs to events. Finally, we obtain an event log on which process mining techniques can be applied.

The remainder of this paper is organized as follows.
Sec.~\ref{sec:relatedWork} reviews some related works.
Sec.~\ref{sec:preliminaries} shows some technical notions on how the SLURM system is implemented and which information is available to eventually form an event log.
Sec.~\ref{sec:approach} explains our approach to apply process mining techniques to the scientific workflows running on SLURM-based HPC clusters.
Sec.~\ref{sec:experiments} introduces some analyses of the event log extracted from the SLURM system of RWTH Aachen University.
Finally, Sec.~\ref{sec:conclusion} concludes the paper.

\section{Related Work}
\label{sec:relatedWork}
Many studies have analyzed HPC behavior starting from data collected about the running jobs. In~\cite{kunz2022hpc}, an extension of miniHPC is proposed, to enable job-level monitoring to interpret anomalous behaviors such as load imbalance, CPU and I/O anomalies, or memory leaks. A framework for monitoring, analyzing, and predicting jobs running on PBS-based job scheduler HPCs is defined in~\cite{DBLP:conf/cluster/PalM20}. The monitoring module captures data about the topology of in-use nodes while a job is running. This provides a deeper understanding of how the job is distributed across the HPCs node network. In~\cite{DBLP:conf/cluster/DietrichWKN20}, a software stack for center-wide and job-aware cluster monitoring to influence node performance is described. 

Process mining techniques have been used to analyze scientific/business workflow logs. In~\cite{DBLP:conf/services/ZengHA11}, a technique to mine scientific workflows based on provenance (i.e., the source of information involved in manipulating artifacts) 
is proposed.
In~\cite{DBLP:journals/tpds/SongCJXYM17}, Scientific Workflow Mining as a Service (SWMaaS) is presented to support both intra-cloud and inter-cloud scientific workflow mining. 

A limitation of~\cite{kunz2022hpc,DBLP:conf/cluster/DietrichWKN20,DBLP:conf/cluster/PalM20} is that they examine the jobs regardless of their interdependencies.
Moreover, in~\cite{DBLP:journals/tpds/SongCJXYM17}, it is assumed that the data source already contains all the necessary information to apply process mining, 
ignoring the situations in which no case notion is defined.
This paper aims to introduce event correlation methods applicable to event data extracted from scientific workflows.

\section{Preliminaries}
\label{sec:preliminaries}

We will focus on analyzing event data from the popular SLURM platform for HPC computing. Hence, in this section, we present some technical notions. To interact with SLURM, we have a set of possible commands. The most essential commands are listed here~\cite{DBLP:conf/jsspp/YooJG03}:

\begin{itemize}
\item \texttt{srun}: runs a single job. We need to create a \texttt{srun} script, which can then be submitted on SLURM for real-time execution.
\item \texttt{sbatch}: submits one or more \texttt{srun} commands for later execution on SLURM.
\item \texttt{squeue}: reports the states of the running jobs. This command helps us to extract a log for process mining purposes.
\end{itemize}

\subsection{Execution of a Single Job on SLURM}
Any script runs on SLURM as a job. As mentioned above, the execution of a job on SLURM could be easily done with \texttt{srun} and \texttt{sbatch} containing one single \texttt{srun} command. Understanding the sequential stages a job undergoes for execution, and the data that can be extracted for each job running in the SLURM queue is valuable~\cite{DBLP:conf/jsspp/YooJG03}.

Typically, jobs pass through several states in the course of their execution. There are a total of 24 possible states for a job, however, three states are most common. In the \emph{PENDING (PD)} state, the job awaits resource allocation; in the \emph{RUNNING (R)} state, it is currently allocated, and in the \emph{COMPLETING (CG)} state, the job is undergoing completion.

The SLURM scheduling queue contains all the information about running jobs. To view this information we use the \texttt{squeue} command. The most important features of the jobs that have been used in our study are listed in Table~\ref{features_table}. These features could be extracted with the \texttt{squeue -o "\%a \%i \%E \%o \%t \%g"} command on the SLURM system. This command shows the list of jobs in the SLURM scheduling queue along with their account, job ID, declared dependency, executed command, status, and project ID information~\cite{DBLP:conf/jsspp/YooJG03}.

\begin{table}[b]
\scriptsize
\begin{center}
\caption{Extracted features of running jobs on SLURM system~\cite{DBLP:conf/jsspp/YooJG03}.}
\begin{tabular}{|m{2cm}|m{9.5cm}|}
\hline
\textbf{Column title} & \textbf{Description} \\ \hline
ACCOUNT & Account associated with the job.\\ \hline
JOBID & An unique value as job identifier. \\ \hline
DEPENDENCY & Specify the dependencies of the job on other jobs. This job will not begin execution until these dependent jobs are complete. In the case of a job that cannot run due to job dependencies never being satisfied, the full original job dependency specification will be reported. A value of \texttt{NULL} implies this job has no dependencies. \\ \hline
COMMAND & The command to be executed. \\ \hline
ST & Jobs typically pass through several states in the course of their execution. The typical states are \emph{PENDING, RUNNING, SUSPENDED, COMPLETING}, and \emph{COMPLETED}. ST is the compact state of the job.  \\ \hline
GROUP & Group name of the job. The project ID is reported as GROUP in SLURM. \\ \hline
\end{tabular}
\label{features_table}
\end{center}
\end{table}

\subsection{Execution of a Sequence of Jobs on SLURM}
To explain how to run a series of jobs (sequence of scripts) on the SLURM workflow system, we will go through an example. Consider a user who wants to run four scripts on SLURM, \emph{pre-processing} as the first one, then \emph{parallel-job1} and \emph{parallel-job2}, which can be executed in parallel but must be executed after the \emph{pre-processing} script, because they need its output. Finally, the \emph{merge} script needs the output of the two parallel jobs for its execution. The user can run this sequence of jobs on SLURM in two ways: either manually (without explicit interdependencies) or automatically (with interdependencies).

\begin{figure}[t]
\centering
\includegraphics[width=0.95\textwidth,page=1]{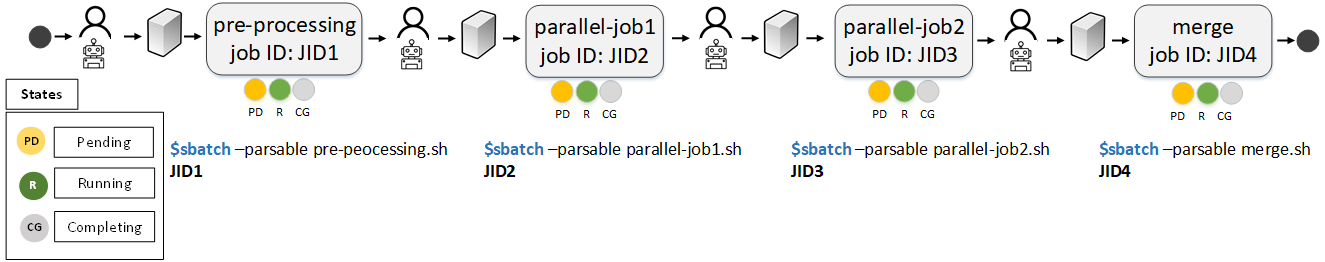}
\caption{Execution of a sequence of jobs without explicit interdependencies.}
\label{without_explicit_interdependencies}
\end{figure}

\textbf{Execution of a Sequence of Jobs without Explicit Interdependencies:} In this case, the user runs the jobs manually---without declaring the inter-dependencies between jobs---and after submitting each job waits for its execution to be completed; then, executes the next job (Fig.~\ref{without_explicit_interdependencies}). In this case, each job is executed as independent, and only the user knows that some of these jobs are logically dependent on each other.

\begin{figure}[ht]
\centering
\includegraphics[width=0.95\textwidth,page=1]{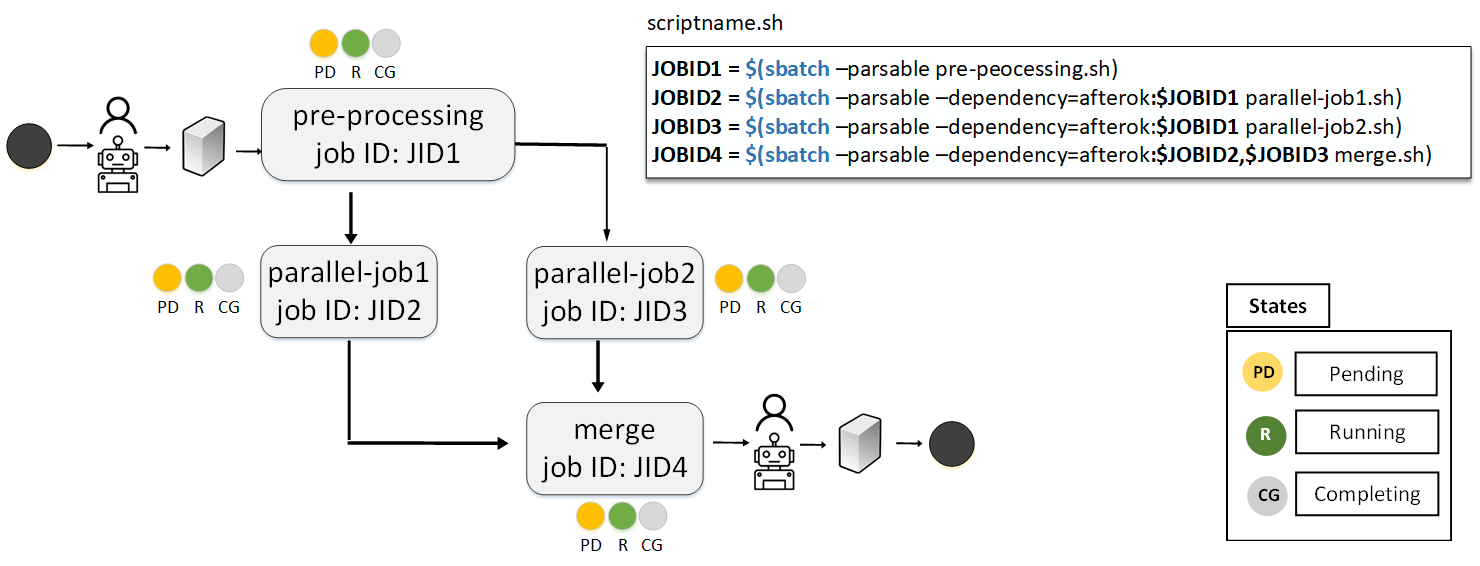}
\caption{Execution of a sequence of jobs with explicit interdependencies.} \label{with_explicit_interdependencies}
\end{figure}

\textbf{Execution of a Sequence of Jobs with Explicit Interdependencies:} In this scenario, the user uses the SLURM dependency management system and submits all jobs at once with correct inter-dependencies on the SLURM system, as shown in Fig.~\ref{with_explicit_interdependencies}. Here, the user uses the \texttt{sbatch} command. This command is used to submit a job script for later execution using the \texttt{--dependency} option. The script typically contains one or more \texttt{srun} commands to launch parallel tasks. In this case, the user does not need to wait for the outputs of a single job, but can wait for the execution of all the tasks and retrieve the final results at completion (Fig.~\ref{with_explicit_interdependencies}).

\begin{table}
\scriptsize
\begin{center}
\caption{Sample SLURM Log}
\begin{tabular}{|m{1.5cm}|m{1cm}|m{2.4cm}|m{2.5cm}|m{1cm}|m{1cm}|m{1.2cm}|}
\hline
ACCOUNT & JOBID & DEPENDENCY & COMMAND & STATE & TIME & GROUP \\ \hline
userA & JID1 & (null) & \texttt{pre-processing.sh} & PD & 13:34:09 & G1\\ \hline
userA & JID2 & \texttt{afterok:JID1 (unfulfilled)} & \texttt{parallel-job1.sh} & PD & 13:34:09 & G1\\ \hline
userA & JID3 & \texttt{afterok:JID1 (unfulfilled)} & \texttt{parallel-job2.sh} & PD & 13:34:09 & G1\\ \hline
userA & JID4 & \texttt{afterok:JID2 (unfulfilled)}, \texttt{afterok:JID3 (unfulfilled)} & \texttt{merge.sh} & PD & 13:34:09 & G1\\ \hline
userA & JID1 & (null) & \texttt{pre-processing.sh} & R & 13:35:00 & G1\\ \hline
userA & JID1 & (null) & \texttt{pre-processing.sh} & CG & 13:49:10 & G1\\ \hline
userA & JID2 & (null) & \texttt{parallel-job1.sh} & R & 13:52:09 & G1\\ \hline
userB & JID5 & (null) & \texttt{Import\_input.sh} & PD & 13:52:32 & G2\\ \hline
userB & JID6 & \texttt{afterok:JID5 (unfulfilled)} & \texttt{Main\_calculation.sh} & PD & 13:52:33 & G2\\ \hline
userB & JID7 & \texttt{afterok:JID6 (unfulfilled)} & \texttt{Export\_output.sh} & PD & 13:52:33 & G2\\ \hline
userA & JID3 & (null) & \texttt{parallel-job2.sh} & R & 13:54:13 & G1\\ \hline
userA & JID3 & (null) & \texttt{parallel-job2.sh} & CG & 14:12:10 & G1\\ \hline
userB & JID5 & (null) & \texttt{Import\_input.sh} & R & 14:12:12 & G2\\ \hline
userA & JID2 & (null) & \texttt{parallel-job1.sh} & CG & 14:38:10 & G1\\ \hline
userB & JID5 & (null) & \texttt{Import\_input.sh} & R & 14:39:32 & G2\\ \hline
userA & JID4 & (null) & \texttt{merge.sh} & R & 14:51:30 & G1\\ \hline
userA & JID4 & (null) & \texttt{merge.sh} & CG & 14:53:30 & G1\\ \hline
userB & JID6 & (null) & \texttt{Main\_calculation.sh} & R & 14:54:12 & G2\\ \hline
userB & JID6 & (null) & \texttt{Main\_calculation.sh} & CG & 14:58:32 & G2\\ \hline
userB & JID7 & (null) & \texttt{Export\_output.sh} & R & 14:59:10 & G2\\ \hline
userB & JID7 & (null) & \texttt{Export\_output.sh} & CG & 15:10:10 & G2\\ \hline
\end{tabular}
\label{slurm_log_with_table}
\end{center}
\end{table}

Table~\ref{slurm_log_with_table} shows the output of the \texttt{squeue} command where the user has declared explicit interdependencies between jobs. As one may see, the \emph{DEPENDENCY} column in the \emph{PENDING} state has a non-empty value. Conversely, when users manually initiate jobs, the \emph{DEPENDENCY} column in the \emph{PENDING} state remains empty, denoted as \emph{(null) values}.

\section{Approach}
\label{sec:approach}

The input of most process mining algorithms is an event log, which contains at least a case, an activity, and a timestamp as attributes for each event. The majority of algorithms presume that the event data is fully accessible and has a clearly defined case notion.

However, we cannot assume that we know the complete historical log, because of privacy issues and the required administrative privileges on the target workflow system. Instead, we aim to observe it for a limited amount of time as described in Sec.~\ref{subsec:registerSLURMevents}. However, this poses technical challenges. For instance, the observation interval may be too long, making it difficult to capture job information during rapid status changes.

Moreover, since the SLURM log does not contain any explicit case notion, in Sec.~\ref{subsec:eventCorrelation} we describe \emph{event correlation} to assign a case to the different events and allow for process mining analyses.

\subsection{Register SLURM events}
\label{subsec:registerSLURMevents}
In order to extract an event log from the system, we perform the following operations periodically (we refer to this as \emph{observing the system}):

\begin{enumerate}
    \item Connect to the access node of the SLURM system
    \item Observe the status (e.g., \emph{PENDING, RUNNING, COMPLETING}) of the current jobs using the \texttt{squeue} SLURM command.
    \item For each of the listed jobs (rows of the log file), one of the following situations occurs:
    \begin{itemize}
        \item The \emph{JOBID} is new: register an event related to the creation of the job.
        \item The \emph{JOBID} already exists, but the status has changed: register an event related to the status change.
        \item The \emph{JOBID} already exists, and the status has not changed: do nothing.
    \end{itemize}
\end{enumerate}

All the features mentioned in Table~\ref{features_table} are recorded for each job. Our log (as illustrated in the sample SLURM log in Table~\ref{slurm_log_with_table}) includes a \emph{TIME} column denoting the event observation time recorded by the script monitoring the SLURM system, and the \emph{COMMAND} values are derived from the executed file path by extracting and using only the file name (the last part of the path).

\subsection{Event Correlation}
\label{subsec:eventCorrelation}

Let us now obtain case IDs from SLURM. We extract a case ID with different techniques, depending on whether the jobs were executed with or without explicit interdependencies.

\textbf{Case ID Extraction with Explicit Interdependencies:}
We utilize this technique when the user has specified the inter-dependencies among jobs. This declaration allows the inclusion of the \emph{DEPENDENCY} column in the extracted log, indicating the jobs on which the current job depends. Note that the \emph{DEPENDENCY} column for the job lists only the dependencies that have not been completed yet. Thus, the \emph{DEPENDENCY} list would be naturally empty for a job that is in the \emph{RUNNING} state.

\begin{figure}
    \centering
    \begin{subtable}{0.6\textwidth}
        \centering
        \tiny
        \begin{tabular}{|m{0.23cm}|m{0.6cm}|m{0.6cm}|m{1.7cm}|m{1.3cm}|m{0.9cm}|m{0.3cm}|m{1cm}|}
            \hline
            & ACC & JID & COMMAND & DEP & TIME & ST & CASE ID\\ \hline
            1 & userA & JID1 & \texttt{pre-processing.sh} & \texttt{(null)} & 13:35:00 & PD & \texttt{JID4321}\\ \hline
            2 & userA & JID2 & \texttt{parallel-job1.sh} & \texttt{afterok:JID1 (unfulfilled)} & 13:52:09 & PD & \texttt{JID4321}\\ \hline
            3 & userA & JID3 & \texttt{parallel-job2.sh} & \texttt{afterok:JID1 (unfulfilled)} & 13:52:11 & PD & \texttt{JID4321}\\ \hline
            4 & userA & JID4 & \texttt{merge.sh} & \texttt{afterok:JID2 (unfulfilled), afterok:JID3 (unfulfilled)} & 13:54:13 & PD & \texttt{JID4321}\\ \hline
            5 & userB & JID5 & \texttt{Import\_input.sh} & \texttt{(null)} & 14:12:12 & PD & JID765\\ \hline
            6 & userB & JID6 & \texttt{Main\_calculation.sh} & \texttt{afterok:JID5 (unfulfilled)} & 14:51:30 & PD & JID765\\ \hline
            7 & userB & JID7 & \texttt{Export\_output.sh} & \texttt{afterok:JID6 (unfulfilled)} & 14:54:12 & PD & JID765\\ \hline
            8 & userA & JID8 & \texttt{pre-processing.sh} & \texttt{(null)} & 14:56:15 & PD & \texttt{JID111098}\\ \hline
            9 & userA & JID9 & \texttt{parallel-job2.sh} & \texttt{afterok:JID8 (unfulfilled)} & 14:59:10 & PD & \texttt{JID111098}\\ \hline
            10 & userA & JID10 & \texttt{parallel-job1.sh} & \texttt{afterok:JID8 (unfulfilled)} & 15:10:10 & PD & \texttt{JID111098}\\ \hline
            11 & userA & JID11 & \texttt{merge.sh} & \texttt{afterok:JID9 (unfulfilled), afterok:JID10 (unfulfilled)} & 16:05:17 & PD & \texttt{JID111098}\\ \hline
        \end{tabular}
        \caption{}\label{jobid_dependency_study_table}
    \end{subtable}
    \hfill
    \begin{subfigure}{0.38\textwidth}
        \centering
        \includegraphics[width=\textwidth,page=1]{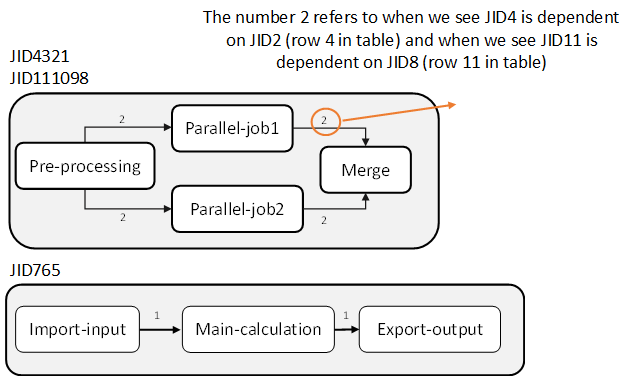}
        \caption{}\label{jobid_dependency_study_model}
    \end{subfigure}
    
    \caption{Case ID extraction with explicit interdependencies by studying \emph{JOBID} and \emph{DEPENDENCY}. ACC stands for ACCOUNT, DEP for DEPENDENCY, JID for JOBID, and ST for STATE. We use these shorter forms to make the table more compact.} \label{jobid_dependency_study}
\end{figure}

To implement this method, a Directed Acyclic Graph (DAG) is generated for each chain of connected jobs in \emph{PENDING} state by utilizing the \emph{JOBID}, and \emph{DEPENDENCY} columns. The vertices are job IDs and the edges show dependent job IDs, and then a unique case ID will be assigned to all of the connected job IDs as shown in Fig.~\ref{jobid_dependency_study_model}. In the table in Fig.~\ref{jobid_dependency_study}, we observe that JID2 and JID3 are dependent on JID1, and JID4 depends on both JID2 and JID3. Based on these dependencies, we assign case ID JID4321, as indicated in the \emph{CASE ID} column of the table. Different cases will be assigned to different discovered connected components. For instance, JID111098 is assigned to another execution of the same chain of commands as JID4321.

\textbf{Case ID Extraction without Explicit Interdependencies:}
\label{without-explicit-interdependencies}
In this case, we do not have explicitly defined job dependencies; therefore, we need to use the attributes at the event level to determine correlations and dependencies between the jobs. We use a combination of the following two attributes in order to define the case identifier:

\begin{itemize}
    \item The \emph{account} executing the job: it is reported as \emph{ACCOUNT} in SLURM.
    \item The \emph{group} of the given job: the project ID is intended to group the jobs belonging to the same project. The status should be empty or default if the user does not call the command with the project ID. The project ID is reported as \emph{GROUP} in SLURM. So, whenever we execute the same scripts several times, the project ID is reported as the same \emph{GROUP} in SLURM. 
\end{itemize}

\begin{figure}
    \centering
    \begin{subtable}{0.56\textwidth}
        \centering
        \tiny
        \begin{tabular}{|m{0.25cm}|m{0.65cm}|m{0.7cm}|m{1.8cm}|m{0.3cm}|m{0.9cm}|m{0.3cm}|m{1cm}|}
            \hline
            & ACC & JID & COMMAND & ST & TIME & G & CASE ID\\ \hline
            1 & userA & JID1 & \texttt{pre-processing.sh} & R & 13:35:00 & G1 & \texttt{userA-G1}\\ \hline
            2 & userA & JID2 & \texttt{parallel-job1.sh} & R & 13:52:09 & G1 & \texttt{userA-G1}\\ \hline
            3 & userC & JID3 & \texttt{pre-processing.sh} & R & 13:52:11 & G1 & \texttt{userC-G1}\\ \hline
            4 & userA & JID4 & \texttt{parallel-job2.sh} & R & 13:54:13 & G1 & \texttt{userA-G1}\\ \hline
            5 & userB & JID5 & \texttt{Import\_input.sh} & R & 14:12:12 & G2 & \texttt{userB-G2}\\ \hline
            6 & userA & JID6 & \texttt{merge.sh} & R & 14:51:30 & G1 & \texttt{userA-G1}\\ \hline
            7 & userB & JID7 & \texttt{Main\_calculation.sh} & R & 14:54:12 & G2 & \texttt{userB-G2}\\ \hline
            8 & userC & JID8 & \texttt{parallel-job2.sh} & R & 14:56:15 & G1 & \texttt{userC-G1}\\ \hline
            9 & userB & JID9 & \texttt{Export\_output.sh} & R & 14:59:10 & G2 & \texttt{userB-G2}\\ \hline
            10 & userC & JID10 & \texttt{parallel-job1.sh} & R & 15:10:10 & G1 & \texttt{userC-G1}\\ \hline
            11 & userC & JID11 & \texttt{merge.sh} & R & 16:05:17 & G1 & \texttt{userC-G1}\\ \hline
        \end{tabular}
        \caption{}\label{account_group_study_table}
    \end{subtable}
    \hfill
    \begin{subfigure}{0.39\textwidth}
        \centering
        \includegraphics[width=\textwidth,page=1]{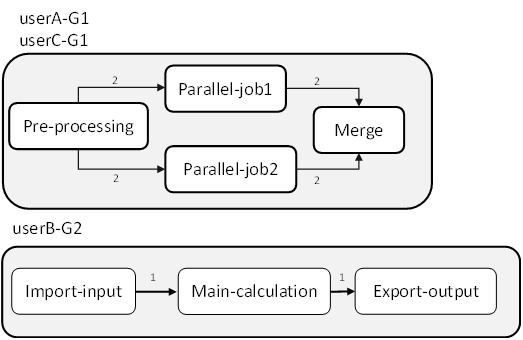}
        \caption{}\label{account_group_study_model}
    \end{subfigure}
    
    \caption{Case ID extraction without explicit interdependencies by studying \emph{ACCOUNT} and \emph{GROUP}. ACC stands for ACCOUNT, JID for JOBID, ST for STATE and G for GROUP. We use these shorter forms to make the table easier to read and understand.} \label{account_group_study}
\end{figure}

We have a many-to-many relationship between accounts and groups. All the jobs executed by an account under a given group are therefore related to the same project. We can use \emph{ACCOUNT-GROUP} as case ID; in this case, we are certain that the jobs executed by the same user under the same project are all collected. In this technique, we generate a unique case ID per each unique combination of \emph{ACCOUNT} and \emph{GROUP}, as shown in Table shown in Fig~\ref{account_group_study_table}. The parallel relationship between \emph{Parallel-job1} and \emph{Parallel-job2} has been discovered based on their occurrence in rows 2, 4 and 8, 10, which show they can be executed in any order.

In this method, we may consider only the account instead of considering the combination of group and account, but the advantage of considering also the group is that the control flow of different projects of the same account is not combined. This technique also has limitations, considering loops where they shouldn't be, due to the inability to differentiate between consecutive executions of the same command related to different experiments. As a result, the precision is significantly reduced, because many different behaviors and command sequences are allowed by the resulting model.

\section{Experiments}
\label{sec:experiments}

\begin{table}[t]
\scriptsize
\begin{center}
\caption{Some event log statistics extracted from the SLURM system. The system was observed in a time interval from 2022-12-07 11:51:45 to 2022-12-09 10:49:07.}
\begin{tabular}{|m{10cm}|m{1cm}|}
\hline
Number of events & 81632 \\ \hline
Number of unique submitted jobs & 17997 \\ \hline
Number of accounts & 123 \\ \hline
Percentage of accounts who submitted jobs with explicit interdependencies& 0.06\% \\ \hline
Percentage of jobs defined with explicit interdependencies & 0.02\% \\ \hline
Average number of allocated CPUs per job & 13.71 \\ \hline
Average amount of allocated RAM per job & 5G \\ \hline
\end{tabular}
\label{table_statistics}
\end{center}
\end{table}

In our previous work, we developed SLURMminer, a tool specifically designed for mining and analyzing process models from SLURM job logs~\cite{DBLP:conf/bpm/SadeghibogarB0A23}. The experiments presented in this paper utilize SLURMminer. SLURMminer was applied to the SLURM system at RWTH Aachen University multiple times, resulting in the extraction of an event log\footnote{For more details on the RWTH HPC cluster, visit: \url{https://help.itc.rwth-aachen.de/service/rhr4fjjutttf/}. A sample event log can be downloaded at: \url{https://www.ocpm.info/hpc_log.csv}.}. Table~\ref{table_statistics} presents key statistics derived from the event log.

Given the variety of research areas (including physics, chemistry, biology, and computer science) and executed purpose-specific scripts, generating a comprehensive process model containing the behavior of all the accounts proved impractical. Instead, we focused on individual account process models. These process models show the scientific workflows executed by a single user/research group. Moreover, the process model is annotated with performance information on the arcs, allowing for the detection of paths with high execution time (bottlenecks), and therefore fulfilling the second goal of finding root causes of performance problems.

To highlight different execution paradigms, we focus on two accounts:
\begin{itemize}
\item \emph{jara0180}: contains computations performed on a funded research project (Molecular dynamics simulations of P2X receptors).
\item \emph{thes1331}: contains scientific experiments performed for an MSc thesis.
\end{itemize}

\begin{figure}[t]
\centering
\begin{subfigure}{0.99\textwidth}
\centering
\includegraphics[width=\textwidth,page=1]{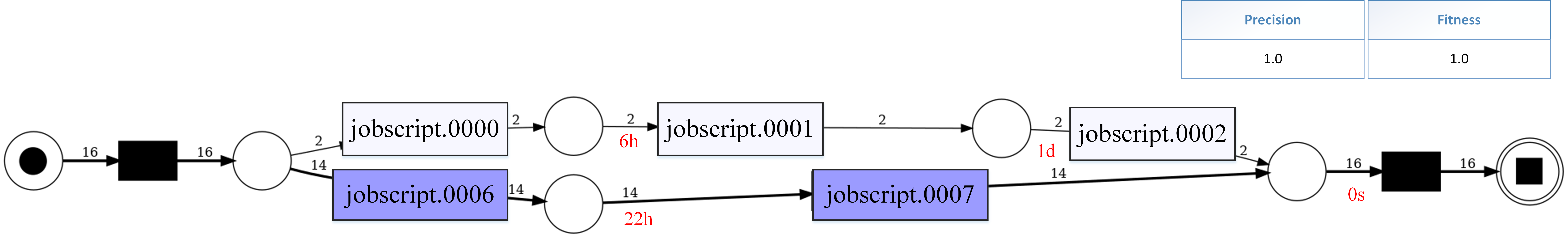}
\caption{}\label{figure11-a}
\vspace{1cm}
\end{subfigure}
\begin{subfigure}{0.99\textwidth}
\centering
\includegraphics[width=\textwidth,page=1]{jara0180_ACCOUNT-GROUP.png}
\caption{}\label{figure11-b}
\end{subfigure}
\caption{The discovered process models for the account \emph{jara0180} considering (a) explicit and (b) implicit interdependencies.} \label{figure11-a-b}
\end{figure}

\begin{figure}[t]
\centering
\begin{subfigure}{0.85\textwidth}
\centering
\includegraphics[width=\textwidth,page=1]{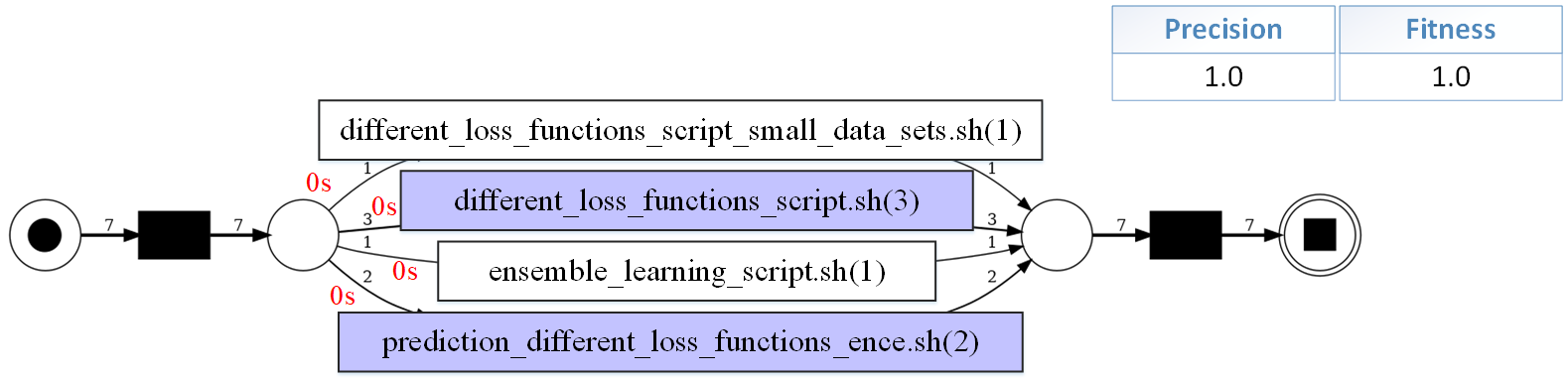}
\caption{}\label{figure12-a}
\vspace{0.5cm}
\end{subfigure}
\begin{subfigure}{0.95\textwidth}
\centering
\includegraphics[width=\textwidth,page=1]{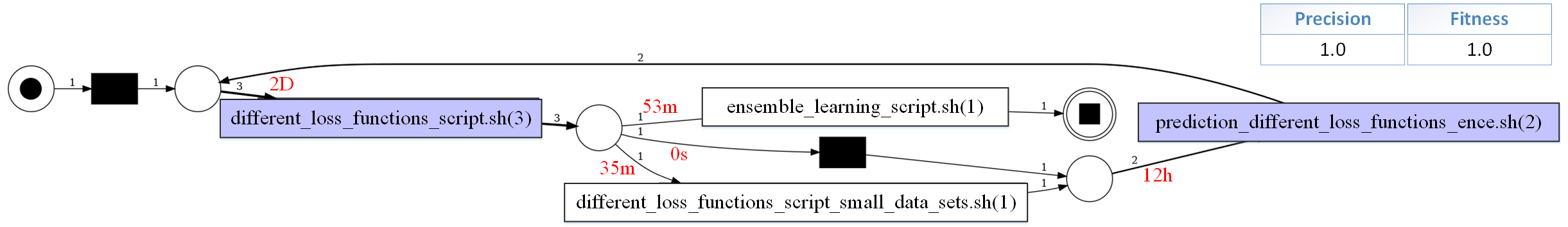}
\caption{}\label{figure12-b}
\end{subfigure}
\caption{The discovered models for the account \emph{thes1331} considering explicit (a) and implicit (b) interdependencies.} \label{figure12-a-b}
\end{figure}

The executions carried out by \emph{jara0180} take advantage of explicit interdependencies (since HPC expertise is involved). Therefore, for \emph{jara0180} we were able to develop a meaningful process model, as depicted in Fig.~\ref{figure11-a}.  In this figure, we observe two distinct chains of commands. The first chain comprises commands from \emph{jobscript\_0000} to \emph{jobscript\_0002} , while the second chain includes the remaining commands related to two different projects and corresponding to 30 distinct cases in the event log. We could still obtain a model from the data (contained in Fig.~\ref{figure11-b}) without considering these interdependencies (considering the \emph{ACCOUNT} and \emph{GROUP} values lead to two distinct cases). However, this model is less precise because it relies solely on the temporal order of command execution, where every event belongs to the same case in the event log.

The executions performed by \emph{thes1331} are defined without explicit interdependencies. The case extraction approach, relying on explicit interdependencies, leads to the assignment of a unique case ID to each execution (seven distinct cases). Consequently, the model depicted in Fig.~\ref{figure12-a} exhibits concurrency among all the executed commands, rendering it highly imprecise. For \emph{thes1331}, it is more appropriate to focus on the models discovered without considering the explicit interdependencies (contained in Fig.~\ref{figure12-b}), which shows the temporal order of execution of the commands. 

The discovered models offer insights by visualizing control flow and execution frequency, aiding users in identifying bottlenecks for improvement. Table~\ref{table_statistics} shows that only a small fraction of HPC users submit jobs with explicit interdependencies, crucial for identifying connected jobs. Without these dependencies, resulting models may be imprecise, with events either belonging to different cases or all belonging to the same case.

\section{Conclusion}
\label{sec:conclusion}

In this paper, we propose an approach to extract and analyze process mining event logs of an HPC system (in particular, we focus on the SLURM system). While this is not the first application of process mining to HPC systems, existing techniques assume the case notion to be well-defined in the data source. This assumption is not satisfied by mainstream systems, and we propose two different case notions (using and not using explicit interdependencies). Moreover, we propose the \emph{SLURMminer} as a tool to connect to the HPC system, extract an event log, and perform a process mining analysis. The analyses allow us to document the execution of scientific workflows for different accounts or research groups utilizing process models that are annotated with performance information (allowing us to detect bottlenecks). Therefore, we address our initial research question: How can process mining be applied to SLURM-based HPC clusters to document workflows and identify execution bottlenecks?

Our event logs are extracted from information that is publicly available in the SLURM system (including the command that is executed and the requested environment, i.e., the number of CPUs, RAM, and disk space required). However, we do not know the detailed content of the commands or have access to more advanced profiling options. This would require collaboration with the specific research groups operating in the HPC systems and availability to modify the execution of scientific workflows to accommodate more detailed process mining analyses.

Our process mining analyses rely on a single account or research group. Since the naming schema of the commands is quite arbitrary, we could not identify shared logical steps (e.g., pre-processing, training of ML model, testing of the model) between different accounts; therefore, we could not produce a generic process model. This is indeed a limitation that could not be tackled without properly naming the commands executed on SLURM and without having insights about the commands.

Overall, our approach succeeds in extracting an event log for process mining purposes from the SLURM HPC system, and we can respond to our basic analytical goals. However, given the arbitrary execution styles and naming conventions, we could not produce more general analyses, which remain as a goal for
future work.

\bibliographystyle{splncs04}
\bibliography{references}

\end{document}